\documentclass[conference]{IEEEtran}
\IEEEoverridecommandlockouts
\usepackage{cite}
\usepackage{amsmath,amssymb,amsfonts}
\usepackage{graphicx}
\usepackage{textcomp}

\def\BibTeX{{\rm B\kern-.05em{\sc i\kern-.025em b}\kern-.08em
    T\kern-.1667em\lower.7ex\hbox{E}\kern-.125emX}}

\usepackage{bm}

\usepackage{algorithm2e}

\usepackage{subcaption}

\newcounter{defcounter}
\makeatletter % <=======================================================
\let\myorg@bibitem\bibitem
\def\bibitem#1#2\par{%
  \@ifundefined{bibitem@#1}{%
    \myorg@bibitem{#1}#2\par
  }{%
    \begingroup
      \color{\csname bibitem@#1\endcsname}%
      \myorg@bibitem{#1}#2\par
    \endgroup
  }%
}
\newcommand*{\bibitem@thecoloredarticle}{blue}  % <==================
\makeatother % <========================================================
\newtheorem{remark}{\textbf{Remark}}

\DeclareMathOperator{\suchthat}{s.t.}

\usepackage[scr=boondox]{mathalpha}
\usepackage{mathtools}

\newcommand{\im}{\hat{\imath}}
\newcommand{\jm}{\hat{\jmath}}

\begin{document}

% Interoperable
% \title{A Cyber-Physical Model for Synergized Power System Optimization}

\title{Cooperative Optimization of Grid-Edge Cyber and Physical Resources for Resilient Power System Operation 
% Network Graph Representation
}

% \author{\IEEEauthorblockN{1\textsuperscript{st} Xiang Huo}
% \and
% \and
% \and
% \and
% \and

\author{Xiang~Huo$^*$, Shining~Sun$^*$, Khandaker Akramul~Haque$^*$, Leen~Al Homoud$^*$, 
Ana~E.~Goulart$^\dagger$,
Katherine R.~Davis$^*$ 
\thanks{$^*$Xiang~Huo, Shining~Sun, Khandaker Akramul~Haque, Leen~Al Homoud, and Katherine R.~Davis are with the Department of Electrical {\&} Computer Engineering, Texas A\&M University, College Station, TX, 77843, USA 
(e-mail: {\texttt{ xiang.huo;sshh2;akramwired;leen.alhomoud;katedavis
@tamu.edu}}). 
}
\thanks{$^{\dagger}$Ana~E.~Goulart is with the Department of Engineering Technology and Industrial Distribution, Texas A\&M University, College Station, TX, 77843, USA 
(e-mail: {\texttt{goulart@tamu.edu}}). 
}
\thanks{This work is supported by NSF Award  2220347 and DOE Award DE-CR0000018.
% The authors would like to acknowledge the
% National Science Foundation under Grant 2220347 and the US Department of Energy under award DE-CR0000018, for supporting this work.
}
}

\maketitle

\begin{abstract}
The cooperative operation of grid-edge 
power and energy resources is crucial to improving the resilience of power systems during contingencies. 
However, given the complex cyber-physical nature of power grids, it is hard to respond timely with limited costs for deploying additional cyber and/or phyiscal resources, such as during a high-impact low-frequency cyber-physical event. 
Therefore, the paper examines the design of cooperative cyber-physical resource optimization solutions to control grid-tied cyber and physical resources. 
% 
% To this end, this paper studies the operation of power systems by controlling both cyber and physical resources. 
% 
First, the operation of a cyber-physical power system is formulated into a constrained optimization problem, including the cyber and physical objectives and constraints.
Then, a bi-level solution is provided to obtain optimal cyber and physical actions, including the reconfiguration of cyber topology (e.g., activation of communication links) in the cyber layer and the control of  physical resources  (e.g., energy storage systems) in the physical layer. 
The developed method improves grid resilience during cyberattacks and can provide guidance on the control of coupled physical side resources. Numerical simulation on a modified IEEE 14-bus system demonstrates the effectiveness of the proposed approach.

% such as power generation and firewall rules

% {\color{blue}

% the cyber 

% especially during system contingencies. 

% graph-based network clustering 

% }

\end{abstract}

\begin{IEEEkeywords}
Cyber-physical power system, communication network, cooperative optimization, grid resilience 
\end{IEEEkeywords}

\vspace{-2mm}
\section{Introduction}

% \cite{surenewable}

% {\color{mygreen}
% 
Power systems are inherently cyber-physical, with frequent and mandatory interactions between cyber and physical resources.
In cyber-physical power systems (CPPSs), the operation of physical devices, such as generators, transformers, and loads, are integrated with the operation of cyber devices, such as routers, switches, and firewalls to continuously
monitor, control, and optimize the grid performance \cite{davis2015cyber}.
Typically, the physical side of a CPPS is  responsible for generating, transmitting, and distributing electric power. 
The physical side resources, such as distributed energy resources, have been proven valuable in providing  ancillary services \cite{li2021learning}, enhancing grid efficiency, and improving grid resilience \cite{huo2024review}. 
% 
% }

% \cite{huo2022two_facet}. 

% The physical objectives can include 

The majority of current work focuses on developing resilient physical power systems \cite{mahzarnia2020review}. However, the physical operation of CPPSs heavily relies on cyber infrastructures that are prone to cyber threats, including all the digital and information technologies used to monitor, communicate, and control the physical assets \cite{fard2023exploration,sahu2022generation,rostami2023reliability}.
To address the cyber threats, Sahu \emph{el al.} \cite{sahu2022generation} develop an automatic firewall configuration tool that can streamline the configuration of firewalls for utilities. 
% 
% In \cite{hasan2016study}, the authors examine cybersecurity planning for resource-constrained smart grid networks, utilizing centrality measurements to rank nodes based on their importance.
% 
To predict cyber risks, Rostami \emph{et al.} \cite{rostami2023reliability} use a Bayesian attack graph to simulate attack paths and a Markov model to illustrate the consequences of attacks.
% enabling the prediction of cyber vulnerability probabilities.
% 
% 
% {\color{mygreen}
Despite having distinct characteristics, ignoring the interconnectivity between cyber and physical layers could lead to failures in power system operation, especially under system contingencies. 
To enhance the resilience of CPPSs, another crucial perspective is to understand and learn the cooperation among grid cyber, physical, and cyber-physical components.
For example, a tri-level optimization model is proposed in \cite{lai2019tri} to mitigate coordinated attacks on CPPSs, which involve physically short-circuiting transmission lines after compromising the communication network of protection relays.
Besides, the development of redundant pathways and backup systems has proven efficacy in restoring operations during cyber-physical attacks \cite{kushal2020decision}. 
% In \cite{kushal2020decision}, a cost-effective distributed generation expansion strategy is developed to evaluate different possible allocations of solar photovoltaics and energy storage systems (ESSs) to load buses during contingencies. 
% 

Therefore, cooperative physical and cyber actions are essential to counteract malicious cyber-physical threats in CPPSs. 
In this paper, we investigate the cooperative optimization of grid-edge resources that interact between system operators and end-users for resilient power system operation. 
The contributions of this paper include:
1) Development of a synergized cyber-physical  optimization framework that can  be used to cooperatively operate cyber and physical resources in CPPSs; 
2) Design a  cyber topology reconfiguration and physical power system restoration strategy; 
and 3) Provide bi-level optimal cyber and physical response actions to enhance grid resilience during synthetic cyber-physical events.

In the rest of this paper, Section \ref{problem_formulation} presents the formulation of a CPPS. 
Section \ref{resilient_operation} designs the resilient cyber and physical system 
operation strategy. 
Section \ref{Simulation} provides numerical simulation results. 
Section \ref{Conclusion} concludes the paper.

\section{Problem Formulation}
\label{problem_formulation}

This section examines the operation of physical and cyber resources within a CPPS  by formulating a constrained optimization problem that incorporates both physical and cyber objectives and constraints.

\subsection{Physical Power System}
\label{section_physical_system}

% The DistFlow branch equations can be written in the real form as \cite{baran1989network}:
% % 
% \begin{subequations}
% \begin{align}
%     & \sum_{k \in \mathcal{C}_j} \mathcal{P}_{jk} = \mathcal{P}_{ij} - P_j  + p_j - r_{ij}\mathcal{I}_{ij}^2, &&\forall j\in \mathcal{N} \label{distflow_1}  \\
%     & \sum_{k \in \mathcal{C}_j} \mathcal{Q}_{jk} = \mathcal{Q}_{ij} - Q_j  + q_j - x_{ij}\mathcal{I}_{ij}^2, &&\forall j\in \mathcal{N} \label{distflow_2}  \\
%     &V_i^2 - V_j^2 = 2(r_{ij} \mathcal{P}_{ij} + x_{ij}\mathcal{Q}_{ij}) \nonumber\\
%     & \qquad\qquad\qquad -(r_{ij}^2 + x_{ij}^2)\mathcal{I}_{ij}^2, &&\forall ij \in \mathcal{E} \label{distflow_3} 
% \end{align}
% \end{subequations}
% where $\mathcal{I}_{ij}^2 = (\mathcal{P}_{ij}^2+\mathcal{Q}_{ij}^2)/V_i^2$. 

% 
% Consider the physical side operation of CPPSs as an optimal power flow (OPF) problem.
% \hspace{0mm}\revsecond{{\bf{REV 2.1}}}{\color{mygreen} 
The physical power system operation is formulated into a standard AC optimal power flow (ACOPF) problem.
% }
The OPF minimizes the system operation costs, i.e., generation costs,  subject to physical operational constraints, i.e.,  power balance, generation limits, and voltage bounds. 
% These OPF problems focus on maximizing the 
% grid benefits with various objective functions. 
% 
The physical power network is described by a connected graph $\mathcal{G}\{\mathcal{N},\mathcal{E}\}$,
where the set $\mathcal{N} = \{0,1,\ldots,n\}$ represents the buses, and the set $\mathcal{E} \subset \mathcal{N} \times \mathcal{N}$ represents the lines. The AC power flow equations can be written as \cite{cain2012history}:
\begin{subequations} 
% \vspace{-4mm}
\label{eq_powerflow_full}
     \begin{align}
       &  P_{ij}   =  (V_i^2 -V_iV_j\cos(\theta_i-\theta_j))g_{ij} \nonumber\\
         &\qquad\qquad\qquad -V_iV_j\sin(\theta_i-\theta_j)b_{ij},&& \forall ij\in \mathcal{E}   \label{eq:pf:P}
\\
 & Q_{ij} = (V_iV_j\cos(\theta_i-\theta_j)-V_i^2)b_{ij}\nonumber \\
         &\qquad\qquad\qquad  -V_iV_j \sin(\theta_i-\theta_j)g_{ij},&&  \forall ij \in \mathcal{E} \\
 & \sum_{j\in \mathcal{N}_i} P_{ij}\, = \sum_{g\in \mathcal{G}_i} P_{i,g}^{\mathrm{gen}} - \sum_{l\in \mathcal{L}_i} P_{i,l}^{\mathrm{load}}, &&  \forall i\in \mathcal{N}  \label{eq:full:p} \\
    &   \sum_{j\in \mathcal{N}_i} Q_{ij} = \sum_{g \in \mathcal{G}_i} Q_{i,g}^{\mathrm{gen}} - \sum_{l\in \mathcal{L}_i} Q_{i,l}^{\mathrm{load}}, &&   \forall i\in \mathcal{N}
     \end{align}
 \end{subequations}
where $P_{ij}$ and $Q_{ij}$ denotes the active and reactive power flow of line $ij$ from bus $i$ to bus $j$, respectively,  
$V_i$ denotes the voltage magnitude of bus $i$,  
$\theta_i$ denotes phase angle at bus $i$, 
$\mathcal{N}_i$ denotes the set of neighbor buses of bus $i$, 
$\mathcal{G}_i$ and $\mathcal{L}_i$ denote the set of generators and loads at bus $i$, respectively, 
$g_{ij}$ and $b_{ij}$ denote the conductance and susceptance of line $ij$, respectively, 
$P_{i,g}^{\mathrm{gen}}$ and  $Q_{i,g}^{\mathrm{gen}}$ denote the active and reactive power output of generator $g$ at bus $i$, respectively, 
$P_{i,l}^{\mathrm{load}}$ and $Q_{i,l}^{\mathrm{load}}$ denote the active and reactive power of load $l$ at node $i$, respectively.

% The network is tree-structured where

% $\mathcal{C}_j$ denote the set of children of bus $j$, 
% and let the line $l_{jk} \in \mathcal{E}$ connect two neighboring nodes, \textit{Node $j$} and \textit{Node $k$}. The active and reactive power flows from \textit{Node i} and \textit{Node j} are represented by $\mathcal{P}_{ij}$ and $\mathcal{Q}_{ij}$, respectively, the resistance and reactance of the line $l_{ij}$ are given by $r_{ij}$ and $x_{ij}$, respectively. Let $P_i$, $Q_i$, $p_i$, and $q_i$ denote the active power consumption, reactive power consumption, active power injection, and reactive power injection to \textit{Node i}, respectively.

\subsubsection{Physical Objectives}

The physical objective minimizes the generator's generation cost, defined as:
\begin{equation} \label{generation_obj}
f_{\mathrm{power}} \coloneqq \sum_{i \in \mathcal{N}} \sum_{g \in \mathcal{G}_{i}}\left(c_{i, g}^{2}\left(P_{i, g}^{\mathrm{gen}}\right)^{2}+c_{i, g}^{1} P_{i, g}^{\mathrm{gen}}+c_{i, g}^{0}\right)
\end{equation}
where $c_{i, g}^{2}$, $c_{i, g}^{1}$, $c_{i, g}^{0}$ are cost parameters of the $g$th generator at bus $i$. 

% The voltage regulation objective can be written into: 
% \begin{equation} \label{voltage_obj}
% f_{\mathrm{vol}} \coloneqq \sum_{i \in \mathcal{N}} \left(c_{i, g}^{2}\left(P_{i, g}^{\mathrm{G}}\right)^{2}+c_{i, g}^{1} P_{i, g}^{\mathrm{G}}+c_{i, g}^{0}\right)
% \end{equation}
% where $c_{i, g}^{2}$, $c_{i, g}^{1}$, $c_{i, g}^{0}$ denote the 

\subsubsection{Physical Constraints}

The physical constraints include the  generator's generation limits, described by: 
\begin{equation} \label{generator_limit}
  \underline{P}_{i,g}  \leq P_{i,g} \leq  \overline{P}_{i,g} , \qquad  \forall i\in \mathcal{N}, g\in \mathcal{G} 
\end{equation}
where $\underline{P}_{i,g}$ and $\overline{P}_{i,g}$
denote the lower and upper bounds of the $g$th generator at bus $i$. 

The voltage constraint can be expressed as: 
\begin{equation} \label{voltage_limit}
\underline{v} V_{0} \leq V_i \leq \overline{v} V_{0}, \qquad \forall i\in \mathcal{N} 
\end{equation}
which requires that the voltage magnitudes of all buses must be constrained within the range of $\left[\underline{v}V_{0}, \bar{v}V_{0}\right]$, $\underline{v}$ and $\overline{v}$ represent the lower and upper bounds, respectively.

\subsubsection{Physical Power System Operation} The physical side power system optimization problem is formulated as:
\begin{subequations} 
% \vspace{-4mm}
\label{physical_opt}
\begin{align}
& 
% {\mathscr{p}(\bm{x})} \coloneqq 
\min_{P_{i, g}^{\mathrm{gen}}, \, Q_{i, g}^{\mathrm{gen}}} \ \ f_{\mathrm{power}} \\
& \: \, \ \ \ \suchthat \ \ \eqref{eq_powerflow_full}, \eqref{generator_limit}, \eqref{voltage_limit}.
% &  &&\bm{x} \in \mathcal{G}.
\end{align}
\end{subequations}

\subsection{Cyber Network}
\label{section_cyber_system}

Cyber (communication) network represents the communication topology among cyber nodes whose connectivity is controlled directly by cyber components, such as routers/firewalls. 
% Based on the router/firewall rules, 
% It is critical to model and control the cyber components to optimize the power grid operation.
% 
The cyber network is represented by a directed graph $\mathcal{G}' \{\mathcal{N}', \mathcal{E}'\}$, where $\mathcal{N}' = \{1, 2, \dots, c\}$ denotes
the set of cyber nodes and $\mathcal{E}' \subset \mathcal{N}' \times \mathcal{N}'$  denotes the set of communication links between cyber nodes. Here, we use nodes and links to distinguish `buses' and `lines' in the physical power system. 
% The cyber topology can capture the connections to the substation remote terminal units and critical physical assets. 
% 
The cyber layer ensures connectivity among critical cyber nodes (e.g., critical cyber nodes  can be defined by vulnerability score) while minimizing the costs of establishing such a communication network.

% is formulated as a flow-based connectivity optimization problem, which

\subsubsection{Cyber Objectives}

The cyber objective minimizes the total cost of deploying cyber resources and activating communication links, defined by:
\begin{equation}
f_{\mathrm{cyber}} \coloneqq \sum_{(\im,\jm) \in \mathcal{E}'} c'_{\im \jm} \cdot y_{\im \jm} + \sum_{\im \in \mathcal{V}'} c'_{\im} \cdot x_{\im}
\end{equation}
where $y_{\im \jm} \in \{0,1\}$ denotes activation status of the communication
link $\im\jm$ between node $\im$ and node $\jm$, $c'_{\im \jm}$ denotes the cost of activating the communication link between nodes $\im$ and $\jm$, $x_{\im} \in \{0,1\}$ denotes the presence of a cyber resource at node $\im$, and $c'_{\im}$ is the hardware/software cost of deploying a cyber resource at node $\im$.

\subsubsection{Cyber Constraints} 
Inspired by the Spanning Tree Protocol (STP) \cite{STP_IEEE}, the cyber communication network configuration is framed as a flow-based topology optimization problem. 
% 
% \revsecond{{\bf{REV 2.2}}}{\color{mygreen}
The STP is designed to function only on a ``tree-like'' network topology, meaning a graph without loops.
% }
In the cyber graph, by assuming an amount of information flows out of a root node to decrease till it reaches the end nodes, the cyber constraints are enforced on flow balance, flow capacity, link activation, resource deployment, and root node flow, respectively.

First, the total flow out of the root node $\im_r$ must equal the number of selected cyber nodes minus one, ensuring that the network forms a spanning tree:
\begin{equation} \label{flow_tree}
\sum_{\jm: (\im_r, \jm) \in \mathcal{E}'} h_{\im_r\jm} = \sum_{\im \in \mathcal{V}'} x_{\im} - 1
\end{equation}
where $h_{\im_r\jm}$ denotes the flow on link $\im_r\jm$. Eq. \eqref{flow_tree} ensures that the flow originates from the root node and connects all active nodes in the cyber network.

Then, the total inflow and outflow of every cyber node except the root node, denoted as $\im \in \mathcal{V}'_{\Bar{\im}_r}$, must satisfy:
\begin{equation} \label{inoutflow_balance}
\sum_{\jm: (\jm, \im) \in \mathcal{E}'} h_{\jm \im} - \sum_{\jm: (\im, \jm) \in \mathcal{E}'} h_{\im\jm} = x_{\im}. \qquad \forall \im \in \mathcal{V}_{\Bar{\im}_r}
\end{equation}
% 
% where $\mathcal{V}'_{\Bar{\im}_r}$ denotes the set of cyber nodes excluding the root node $\im_r$.
% 
Eq. \eqref{inoutflow_balance} ensures that net flows at these nodes must equal the binary indicator $x_{\im}$ that represents the nodal status, i.e., active (1) and inactive (0).

The flow on each link is constrained further by an activation level of the communication link, defined by:
\begin{equation} \label{flow_limit}
h_{\im\jm} \leq M \cdot y_{\im\jm}, \qquad  \forall \im\jm \in \mathcal{E}'
\end{equation}
where $M$ is a large constant representing the maximum possible flow on any edge, e.g., $M$ is the number of nodes.

Moreover, a communication link can be activated only if both endpoints are active, meaning that cyber resources are deployed at both nodes.
This is described by the following edge activation and resource deployment constraints:
\begin{equation} \label{activation_require} 
y_{\im\jm} \leq \min \{x_{\im},x_{\jm}\}, \qquad \forall \im\jm \in \mathcal{E}'
\end{equation}
which ensures that the flow on link $\im\jm$ is only allowed if cyber resources are deployed at both nodes $\im$ and $\jm$.

\subsubsection{Cyber Resource Optimization}

To summarize, the flow-based cyber topology  optimization problem is formulated as:
\begin{subequations}  
\vspace{-4mm}
\label{cyber_opt}
\begin{align} 
& \min_{y_{\im \jm}, \, x_{\im}} \ \ f_{\mathrm{cyber}} \label{11a} \\
& \: \, \ \suchthat \ \ \eqref{flow_tree}, \eqref{inoutflow_balance}, \eqref{flow_limit}, \eqref{activation_require}. \label{11b}
\end{align} 
\end{subequations}
% 
% \begin{equation}
% \min
% \end{equation}
% subject to:
% \begin{equation}
% \sum_{j: (i, j) \in E} f_{ij} - \sum_{j: (j, i) \in E} f_{ji} = x_i \quad \forall i \in V, i \neq 1
% \end{equation}
% \begin{equation}
% f_{ij} \leq M \cdot y_{ij} \quad \forall (i, j) \in E
% \end{equation}
% \begin{equation}
% y_{ij} \leq x_i \quad \text{and} \quad y_{ij} \leq x_j \quad \forall (i, j) \in E
% \end{equation}
% \begin{equation}
% \sum_{j: (1, j) \in E} f_{1j} = \sum_{i \in V} x_i - 1
% \end{equation}
% \begin{equation}
% 0 \leq y_{ij} \leq 1 \quad \forall (i, j) \in E
% \end{equation}
% \begin{equation}
% 0 \leq x_i \leq 1 \quad \forall i \in V
% \end{equation}

Therefore, problem \eqref{cyber_opt} takes the form of a mixed integer linear programming problem that can be solved by existing solvers. 
The flow-based approach ensures that any set of cyber nodes (such as critical nodes) are fully connected with minimized total costs on communication link activation and cyber resource deployment. 
Moreover, it can prioritize communication routes with less transmission delay by decreasing their communication link activation costs, which can improve the communication efficiency of the cyber network.

% \subsubsection{Linearization}

% By relaxing the decision variables $y_{ij}$ and $x_i$ to continuous values, the problem can be solved in  linear optimization with an approximation of the optimal solution. 
% % 
% Specifically, we let $\tilde{y}_{ij} \in [0, 1]$ denote a continuous decision variable that represents the activation level of the communication link between nodes $i$ and $j$, $\tilde{x}_i \in [0, 1]$ denote the level of deployment of a cyber resource at node $i$, and $\tilde{f}_{ij} \geq 0$ denotes the continuous flow variable representing the flow from node $i$ to node $j$ over the communication link $(i, j)$. 

% Subsequently, the flow-based cyber resource optimization problem can be written into a linear form.  
% of: 

% 5. \textbf{Bounds on Variables:}
% \begin{equation}
% 0 \leq y_{ij} \leq 1 \quad \forall (i, j) \in E
% \end{equation}
% \begin{equation}
% 0 \leq x_i \leq 1 \quad \forall i \in V
% \end{equation}

% {\color{mygreen}
\begin{remark} \label{remark1}
The presence of a cyber node can be represented by a router at a substation, and the router communicates with other nodes and the control center via communication links. 
% 
% STP can help manage Ethernet networks with redundant paths, ensuring they remain loop-free while retaining backup paths for resilience.
% 
The cost of building a communication link (e.g., Ethernet path) depends on the bandwidth of each Ethernet link, which can be manually adjusted to find the optimal cyber topology.
% when activating a communication link.
% 
In practice, the cost of building a communication link can also reflect the time delay requirements. 
\hfill $\square$ 
\end{remark}
% }

\section{Resilient Cyber-Physical Power System Operation}
\label{resilient_operation}

In this section, we probe into the coupled operation of cyber-physical resources to enhance grid resilience, then develop a constrained cyber-physical resilience optimization problem with cooperative bi-level solutions.

\subsection{Cyber-Physical Resilience}

Cyber resilience is essential for operating CPPSs, as it enables the rapid reconfiguration of cyber topology between cyber nodes during a system contingency.
We aim to first enhance cyber resilience on the cyber layer during a cyberattack, and then study the coupled cyber and physical actions to optimize the impacted physical power system operation.
In specific, the cyber resilience is achieved by dynamically updating the cyber topology to 1) ensure the interconnection on a set of critical cyber nodes, and 2) interact cyber nodes with the physical layer to adaptively isolate the compromised cyber components.
Regarding cyberattacks, we take the example of combined cyber-physical attacks that can lead to the isolation of a cyber node and the dysfunction of the generators.
% within its communication range. 
% 

Assume a set of critical cyber nodes denoted as $\mathcal{K} = \{\im_{\tilde{1}}, \ldots, \im_{\tilde{k}} \}$, e.g., nodes responsible for controlling generator buses.
During a cyberattack (e.g., false data injection), a corrupted cyber resource $\im_{\tilde{c}} \in \mathcal{K}$ needs to be isolated and replaced.
To this end, we utilize a set of neighboring cyber resources $\mathcal{M}_{\im_{\tilde{c}}}$ of node $\im_{\tilde{c}}$, as potential candidates for reconstructing communication topology after the cyber attack.
% 
% 
% \revfirst{{\bf{REV 1.2}}}{\color{blue} 
Importantly, the selection of neighboring cyber nodes has the benefits of: 1) providing faster control of physical resources with closer geographical distances; and 2) facilitating the reconstruction of cyber communication links with less rerouting costs.
% }
% 
%
Therefore, the cyber reconfiguration guides the control of physical resources (e.g., flexible loads and backup ESSs) to quickly restore the power system to optimal operation states. 

\vspace{-1mm}
\subsection{Cyber-Physical Couplings}

\begin{figure}[!htb]
% \vspace*{-2mm}
    \centering
    \includegraphics[width=0.45\textwidth, trim = 0mm 0mm 0mm 0mm, clip]{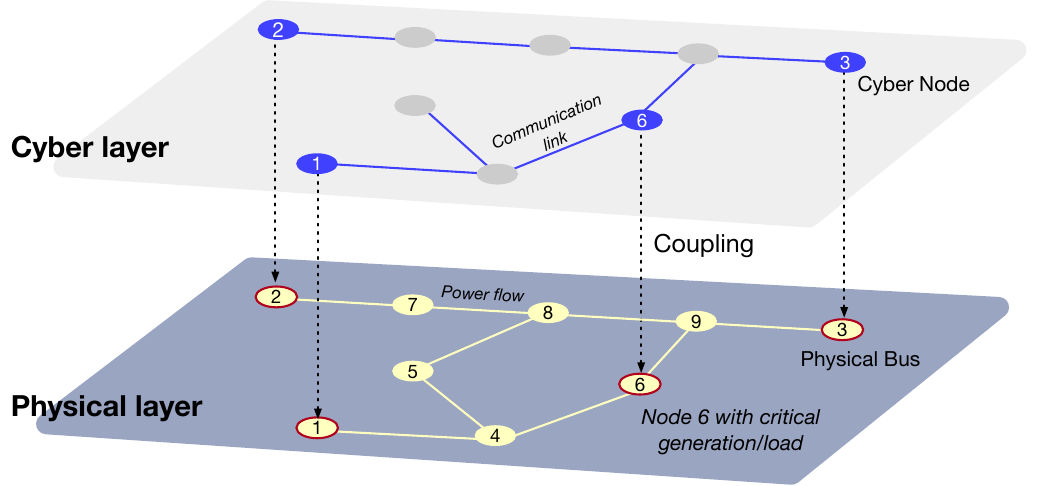}
    % \vspace*{-3mm}
    \caption{A depiction of cyber and physical topologies for CPPSs, with four critical cyber nodes and nine physical buses.}
    \vspace*{-2mm}
    \label{Fig_two_layer_CPS}
\end{figure}

During a cyberattack, the change in the cyber topology could impact both ongoing communication network traffic and the operation of the physical assets. 
Cooperative cyber and physical actions from both cyber and physical layers are needed to achieve resilient grid operations. 
We build the cyber layer by extracting the critical physical buses as cyber nodes to show the cooperative actions from the cyber and physical layers. 
For example, Fig. \ref{Fig_two_layer_CPS} presents the CPPS that is built based on the WECC 9-bus test case \cite{liu2015analyzing}.
% \cite{WSCC_9bus,liu2015analyzing}. 
% 

% \hspace{0mm}\revfirst{{\bf{REV 1.1}}}{\color{blue} 
% 
Though all physical layer nodes can be mapped to the cyber layer, at a minimum, the selection of critical cyber nodes should ensure the direct control of critical physical assets (e.g., generators). 
This allows lower costs on cyber resource deployment and flexibility on reconfiguring the cyber topology.
% }
As shown in Fig. \ref{Fig_two_layer_CPS}, 
% we map all the physical buses in the physical layer into cyber nodes in the cyber layer.
% \revfirst{{\bf{REV 1.1}}}{\color{blue} 
% Then, 
a set of predefined critical cyber nodes $K = \{1,2,3,6\}$ must be active and connected when constructing the cyber topology.
% }
% 
%
Suppose cyber Node 6 is of high resilience priority due to the control of critical generation/load.
On the compromise of Node 6, the cyber decision variable $\tilde{x}_{6}$ is used to decide whether to deploy another cyber resource and control of physical resources at Bus 6. 
For example, deploying an additional cyber resource while enabling a backup physical resource, such as a backup energy storage system (ESS), we have the cost objective of:
\begin{equation} \label{cps_obj}
  f_{\mathrm{res}} (\tilde{x}_{6}, \bm{P}_6^\mathrm{ess}) \coloneqq   \tilde{c}'_{6} \cdot \tilde{x}_{6}  + (c_6^\mathrm{ess} + f_{\mathrm{ESS}}(\bm{P}_6^\mathrm{ess})) \cdot \tilde{x}_{6}
\end{equation}
where $c_6^\mathrm{ess}$ denotes the startup cost of the backup ESS at Node 6, $\bm{P}_6^\mathrm{ess} \in \mathbb{R}^{T}$ denotes the charging/discharging profiles of the ESS across $T$ time intervals, and $f_{\mathrm{ESS}} (\cdot)$ can denote the ESS objective, e.g., battery degradation cost by $f_{\mathrm{ESS}} \coloneqq \| \bm{P}_6^\mathrm{ess} \|_2^2$.  
In return, the participation of a new ESS at the physical layer would require rescheduling of the physical side operation, enforcing new constraints on the ESS by: 
% 
% \begin{subequations} \label{ESS_individual_constraint}
% \begin{align}
%   \underline{P}_i^\mathrm{ess}  &\leq P_{i,t}^\mathrm{ess} \leq \overline{P}_i^\mathrm{ess}  \\
%  \underline{e}_{i} &\leq e_{i, t}^{\mathrm{ess}} \leq \overline{e}_{i}
% \end{align}
% \end{subequations}
\begin{equation} \label{ESS_individual_constraint}
% \begin{align}
  \underline{P}_i^\mathrm{ess}  \leq P_{i,t}^\mathrm{ess} \leq \overline{P}_i^\mathrm{ess}, ~   
 \underline{e}_{i} \leq e_{i, t}^{\mathrm{ess}} \leq \overline{e}_{i}
% \end{align}
\end{equation}
where $e_{i, t}^{\mathrm{ess}}$ denotes the energy of the ESS at node $i$ at time $t$, 
$\underline{P}_i^\mathrm{ess}$ and $\overline{P}_i^\mathrm{ess}$ denote its lower and upper charging power limits, respectively, 
and $\underline{e}_{i}$ and $\overline{e}_{i}$ denote its lower and upper energy bounds, respectively. 
% 
% Eq. \eqref{cps_obj} bridges the reconfiguration of cyber and physical networks to help evaluate the costs of utilizing cyber and physical resources during planning or response. 
% \vspace{-2mm}

% Note this could be extended to multi-period optimization problem with the attack period to adjust the ESS status, otherwise it will be turn off, consider it in the future work. 

\subsection{Cooperative Cyber-Physical Optimization}

To ensure the optimal management of cyber and physical resources during power system contingencies, we develop the following resilient cyber-physical optimization model:
% \mathversion{setB}%
% \begin{pequation} \label{CPS_opt}
% \begin{aligned}
% \min_{\bm{x},\bm{y}} \quad & {\mathscr{c}(\bm{x})} + {\mathscr{p}(\bm{y})} + {\mathscr{r}(\bm{x},\bm{y})} \\
% \textrm{s.t.} \quad & \bm{x}_{i} \in \mathcal{X}_{i}, i=1,2, \ldots, \hat{c}\\
% & \bm{y}_{j} \in \mathcal{Y}_{j}, j=1,2, \ldots, \hat{p}\\
%   &\bm{x}, \bm{y} \in \mathcal{G} 
%     % &\bm{y} \in \mathcal{G} 
% \end{aligned}
% \end{pequation}
% 
% \begin{subequations} \label{CPS_opt}
% \begin{align}
% & 
% % {\mathscr{c}(\bm{x})} \coloneqq
% \min_{y_{\im \jm},x_{\im}} \ \ \alpha_1 f_{\mathrm{cyber}}  + \alpha_2 f_{\mathrm{power}} + \alpha_3 f_{\mathrm{resilience}} \\
% & \: \, \suchthat \ \ \eqref{inoutflow_balance}, \eqref{flow_limit}, \eqref{activation_require}, \eqref{flow_tree}.
% % &  &&\bm{x} \in \mathcal{G}.
% \end{align}
% \end{subequations}
% 
\begin{subequations} \label{CPS_bilevel_opt}
\begin{align}
% &  \notag \\
\textbf{upper} ~~ &  \min_{\mathscr{c}} \ \ \alpha_1 f_{\mathrm{cyber}} \label{13a} \\
& \: \, \suchthat \ \ \eqref{flow_tree}, \eqref{inoutflow_balance}, \eqref{flow_limit}, \eqref{activation_require}. \label{13b} \\
% & \textbf{Lower-level (Power and Resilience):} \notag \\
\textbf{lower}  ~~ & \min_{\mathscr{p}, \mathscr{r}} \ \ \alpha_2 f_{\mathrm{power}} + \alpha_3 f_{\mathrm{res}} \label{13c} \\
& \: \, \suchthat \ \ \eqref{eq_powerflow_full}, \eqref{generator_limit}, \eqref{voltage_limit}, \eqref{ESS_individual_constraint}. \label{13d}
\end{align}
\end{subequations}
% 
% \begin{pequation} \label{CPS_opt}
% \begin{aligned}
% \min_{\bm{x},\bm{y}} \quad & {\mathscr{c}(\bm{x})} + {\mathscr{p}(\bm{y})} + {\mathscr{r}(\bm{x},\bm{y})} \\
% \textrm{s.t.} \quad & \bm{x}_{i} \in \mathcal{X}_{i}, i=1,2, \ldots, \hat{c}\\
% & \bm{y}_{j} \in \mathcal{Y}_{j}, j=1,2, \ldots, \hat{p}\\
%   &\bm{x}, \bm{y} \in \mathcal{G} 
%     % &\bm{y} \in \mathcal{G} 
% \end{aligned}
% \end{pequation}
% $\mathscr{c}(\bm{x})$, $\mathscr{p}(\bm{x})$, and $\mathscr{r}(\bm{x},\bm{y})$ denote the cyber, physical, and resilience objectives, respectively. 
where 
$\mathscr{c} = \{y_{\im \jm}, \, x_{\im} \}$, 
$\mathscr{p} = \{\bm{P}_{i, g}^{\mathrm{gen}}, \, \bm{Q}_{i, g}^{\mathrm{gen}}\}$,
and $\mathscr{r} = \{ \tilde{x}_{i}, \bm{P}_i^\mathrm{ess} \}$ denote the sets of cyber, physical, and resilience variables, respectively, and $\alpha_1$, $\alpha_1$, and $\alpha_3$ are adjustable balancing coefficients. 
Problem \eqref{CPS_bilevel_opt} is framed as a bi-level optimization problem that incorporates both the cyber (upper) and cyber-physical (lower) layers into resilient CPPS operation.  
% 
% The optimal solutions assist in reconfiguring the cyber graph topology, controlling power generators  outputs, and managing backup generations/loads. 

% and 
% \begin{pequation} \label{p2}
% \begin{aligned}
% \min_{\bm{x},\bm{y}} \quad & {\sum_{i=1}^{\hat{c}}c_i(\bm{x}_i)} + {\sum_{j=1}^{\hat{p}}p_j(\bm{y}_j)} + r(\bm{x},\bm{y}) \\
% \textrm{s.t.} \quad & \bm{x}_{i} \in \mathcal{X}_{i}, i=1,2, \ldots, \hat{c}\\
% & \bm{y}_{j} \in \mathcal{Y}_{j}, j=1,2, \ldots, \hat{p}\\
%   &\bm{x}, \bm{y} \in \mathcal{G} 
%     % &\bm{y} \in \mathcal{G} 
% \end{aligned}
% \end{pequation}where 

% Apart from providing ancillary services, achieving resilience is also another essential objective that is embedded into the problem formulation. 

% \subsection{Cyber Physical Optimal Response}
% The Bi-level optimization solution consists of two may steps in the cyber and physical layer

% Alternating layer optimization of CPPS. 
To deploy the cooperative cyber-physical optimization model in \eqref{CPS_bilevel_opt}, we propose an adaptive cyber and physical optimization algorithm shown in Algorithm \ref{alg_1}.
\RestyleAlgo{ruled}
\begin{algorithm}[h]
\caption{Adaptive cyber and physical resource optimization.}
	\label{alg_1}

{\bf Input:} Cyber cost coefficients 
 $c'_{\im \jm}$, 
 % $\forall \im\jm \in \mathcal{E}'$, 
  $c'_{\im}$, $\tilde{c}'_{\im}$, physical cost coefficients $c_{i, g}^{2}$, $c_{i, g}^{1}$, $c_{i, g}^{0}$, $c_i^\mathrm{ess}$, balancing coefficients $\alpha_1$, $\alpha_1$, $\alpha_3$, power system parameters. 
	
{\bf Initialize:} Solve \eqref{cyber_opt} to initialize the cyber topology\; 
% 
% \For{iteration $t=0, 1, \ldots, T-1$}
	% {
		% Each agent collects $N$ Markovian samples, computes its corresponding local importance sampling ratio and broadcasts over the network. \\
	% }	
	% \For{iteration $t=T, T+1, \ldots, T+T'-1$}
	% {	
 
        \uIf{Cyberattack is False}{
            % Monitor the cyber communication network\;
          % Solve \eqref{physical_opt} to update physical power operation
          Solve \eqref{physical_opt} to 
control the physical resources based on the cyber topology\;
        }
        \ElseIf{Cyberattack is True}{
            Isolate the corrupted cyber node $\im_{\tilde{c}}$\;
            % Conduct the flow-based cyber topology reconstruction\;
            	\For{Cyber sources $m\in\mathcal{M}_{\im_{\tilde{c}}}$ in parallel}{
             Update the list of critical cyber nodes by replacing  $\im_{\tilde{c}}$ with $m$\;
			Solve \eqref{13a} and \eqref{13b} to find minimal cyber costs with $\tilde{x}_{6}$\;

		}
     	Output updated cyber topology\;
  Solve \eqref{13c} and \eqref{13d} to obtain
   physical actions in response to the cyberattack\;   
        }
        % \Else{
        %     Perform default action\;
        % }
	% }
	\textbf{Output:} $\mathscr{c},\mathscr{p},\mathscr{r}$.
    % \vspace{-2mm}
\end{algorithm}
Algorithm \ref{alg_1} is suitable for cyber topology planning and providing response after cyber and/or physical attacks. 
% 
% \revfirst{{\bf{REV 1.1}}}{\color{blue} 
In response to a cyber attack, 
% available cyber nodes are candidates that can be activated in the re-optimization. T
the re-optimization allows cyber layer adaptively reroute the communication traffic by isolating the compromised cyber node(s) and updating the list of critical cyber nodes with minimal cyber costs.  
% }
% When a cyberattack happens, 
% 
Afterwards, the physical power system is adjusted to utilize backup ESSs to return to an optimal operation status.

\begin{remark} \label{remark1}
The cyber and physical layers can couple through both cooperative objectives (e.g., resilience goal) and competitive objectives (e.g., resource allocation cost).  
Problem \eqref{CPS_bilevel_opt} decouples the cyber and physical operation by designing a bi-level optimization problem. 
% \hspace{0mm}\revsecond{{\bf{REV 2.3}}}{\color{mygreen} 
Note that the bi-level optimization problem doesnot ensure global optimality across both cyber and physical levels. 
Instead, it ensures optimality at both local cyber and physical levels while providing fast combined cyber and physical actions during cyber-physical contingencies. 
In this way, the bi-level solution also  exemplifies the cyber-physical interactions through the allocation and control of cyber and physical resources.
% }
% at the cyber layer and the allocation of physical energy resources at the physical layer. 
%  
% The proliferation of grid-tied energy devices 
% 
% operation of a ESS that is coupled through both cyber and physical networks. 
% The proliferation of grid-tied energy devices 
% s
% , congestion in the power grid can affect the requirements for communication paths
% 
 \hfill $\square$ 
 % \qed
\end{remark}

\vspace{-1mm}

\section{Numerical Simulation}
\label{Simulation}

% We first introduce the simulation settings for the physical and cyber systems, respectively. 
% 
% Then, we show results of cooperatively optimizing physical and cyber resources for power system operation. 
% 
% \subsection{Physical Network Setup}
% 
In the physical layer, consider the day-ahead power dispatch problem consisting of $T = 12$ time steps with 2-hour intervals. 
The IEEE 14-bus test case is used as the test system, including 5 generators and loads connected to 11 buses (other than the slack bus and PV buses) \cite{IEEE_14bus,Matpower_14bus}. 
Bus 0 serves as the slack bus and maintains a normalized voltage magnitude of $V_0$.
The upper and lower voltage magnitudes are set to be $\overline{V} = 1.06 V_0$ and $\underline{V} = 0.94 V_0$, respectively. 
The generator's cost coefficients $c_{i, g}^{2}$, $c_{i, g}^{1}$, $c_{i, g}^{0}$ are set based on \cite{Matpower_14bus}. 
% 
% \subsection{Cyber Network Setup}
% 
In the cyber layer, we map all physical buses into the cyber layer to obtain a potential list of cyber nodes  based on the physical network topology. 
Then, the communication network consists of a single control center and certain substations, where the routers at the substation are denoted as critical cyber nodes that can control the cyber and physical assets within the area \cite{narimani2021generalized}. 
% 
% 
% The set of generator buses  $\{1,2,3,6,8 \}$ are mapped into the cyber layer and selected to be critical cyber nodes. 
A set of generator nodes $\{1,2,3,6,8\}$ is initiated as critical cyber nodes that need to be connected. Node 1 is assumed to be the control center. 
All critical cyber nodes should be able to communicate with others. 
% by activating a communication link.  
% 
% Alternatively, the cyber communication network can be constructed given the criticalness of the physical nodes. 
% 
% 
% 
% 
The cost coefficient for activating a cyber communication link can be referenced to the IEEE cost of Ethernet paths for the spanning tree \cite{STP_IEEE}. We use solver IPOPT \cite{biegler2009large} to solve the optimization problem.
% 
% To give a clear understanding of the connections between the cyber and physical layers, we consider the equal criticalness of placing an additional neighboring cyber resource. 
% 

\begin{figure}[!htb]
  \vspace{-1mm}
    \centering
    \begin{subfigure}[t]{0.23\textwidth}
        \centering
        \includegraphics[width=\textwidth, trim = 0mm 0mm 0mm 0mm, clip]{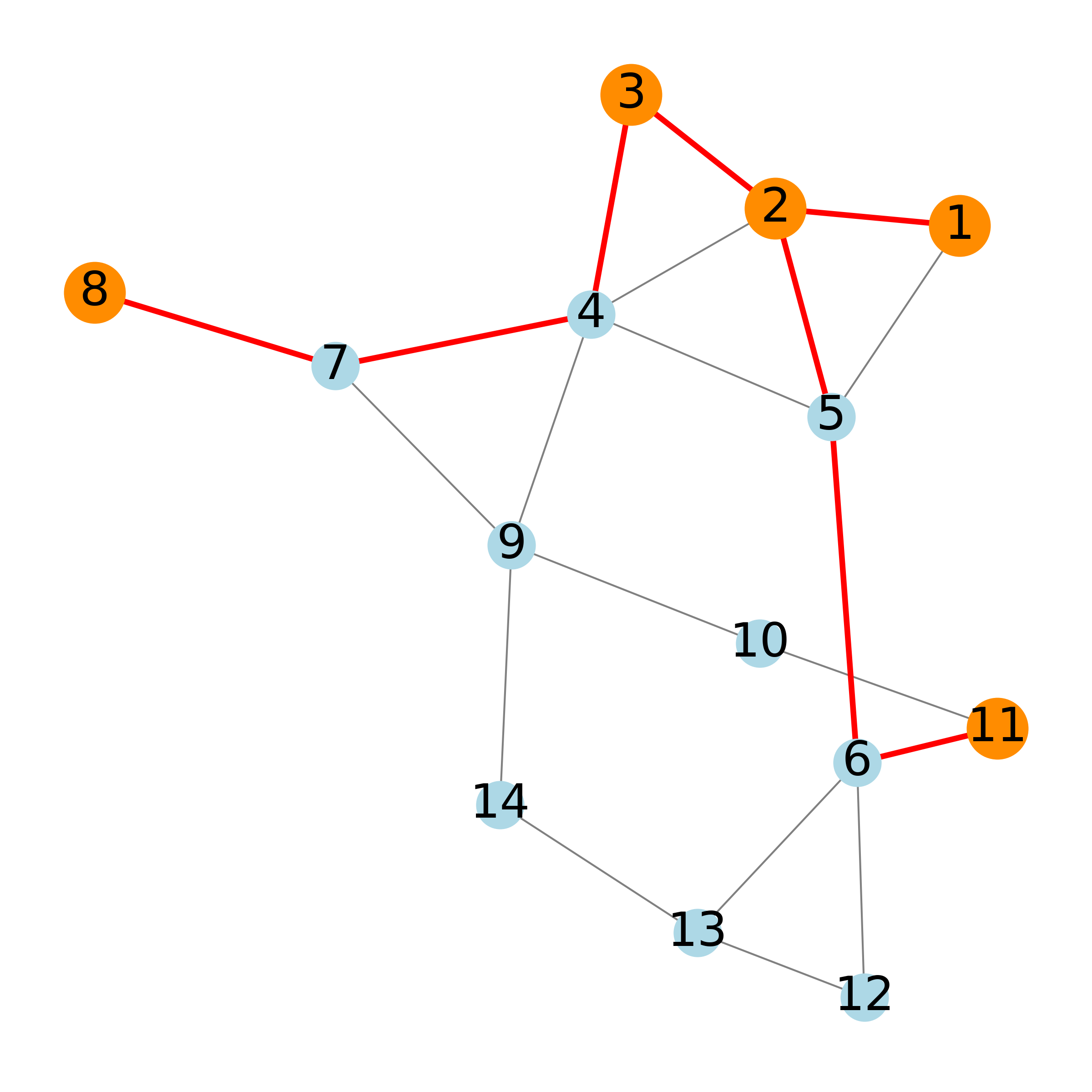}
        \caption{Original cyber topology with Node 1 as the root node (before attack).}
        \label{cyber_network_root8}
    \end{subfigure}
    \hfill
   % \hspace{5mm}
    \begin{subfigure}[t]{0.23\textwidth}
        \centering
        \includegraphics[width=\textwidth, trim = 0mm 0mm 0mm 0mm, clip]{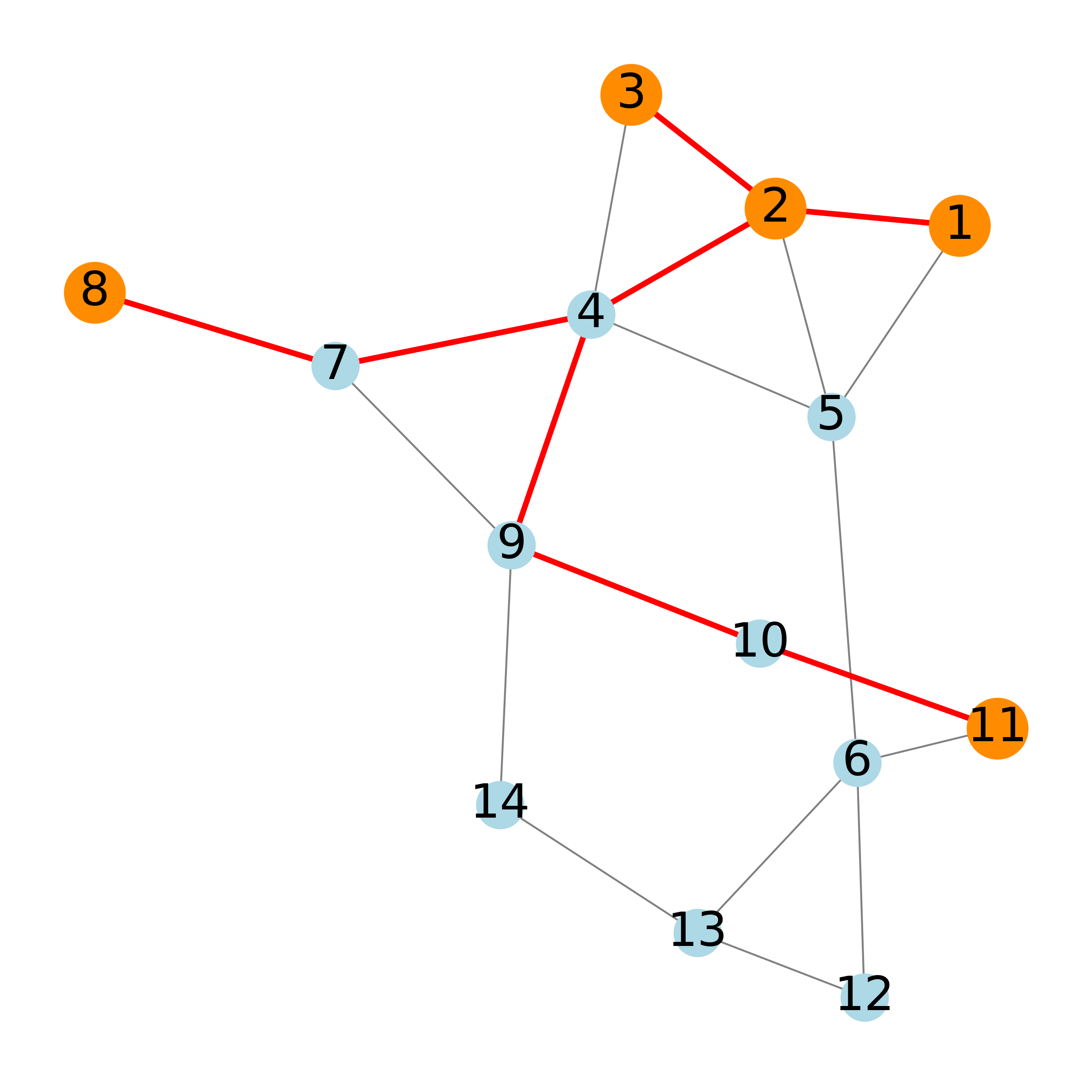}
        \caption{Rerouted cyber topology (after attack).}
        \label{cyber_network_root6}
    \end{subfigure}
    \caption{Construction of cyber communication networks (black solid lines denote the physical power lines, critical cyber nodes are colored in orange).}
    \vspace{-6mm}
    \label{cyber_network}
\end{figure}

\begin{figure}[!htb]
\vspace*{-2mm}
    \centering
    \includegraphics[width=0.45\textwidth, trim = 0mm 0mm 0mm 0mm, clip]{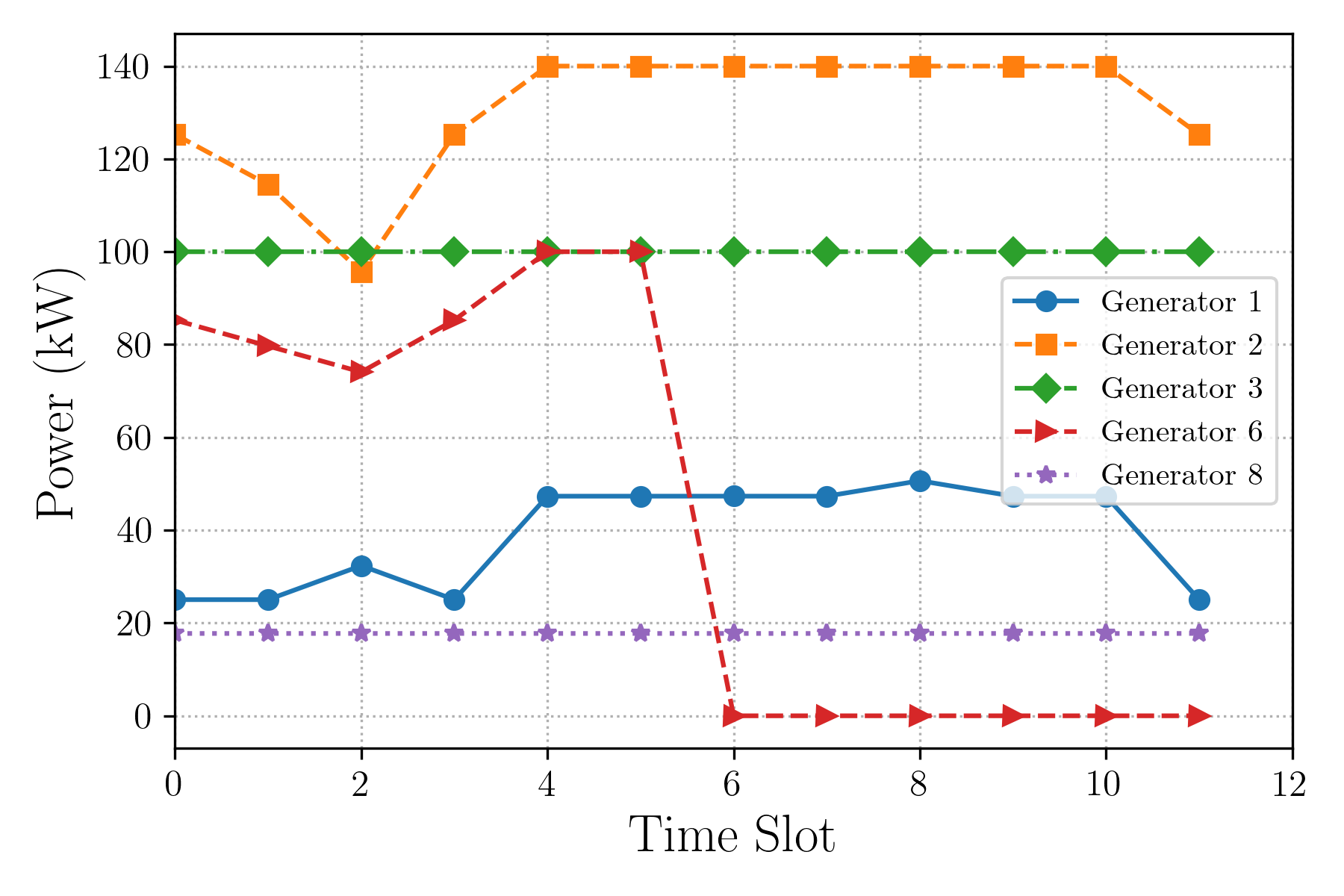}
    \vspace*{-1mm}
    \caption{Active power generation of all generators after the  cyber-physical attack.}
    \vspace*{-2mm}
    \label{all_generator_after_attack}
\end{figure}

\begin{figure}[!htb]
\vspace*{-2mm}
    \centering
    \includegraphics[width=0.47\textwidth, trim = 0mm 0mm 0mm 0mm, clip]{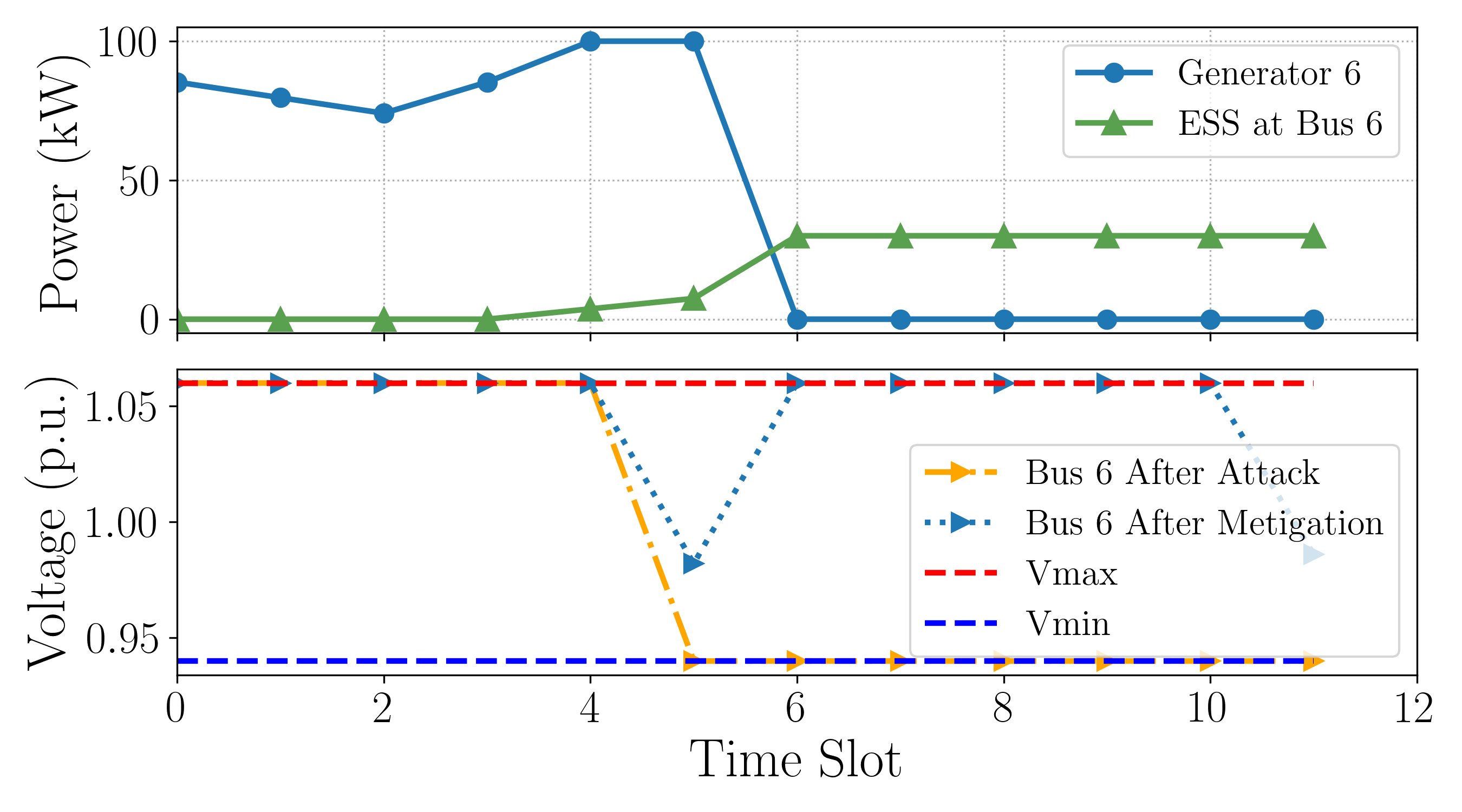}
    % \vspace*{-3mm}
    \caption{Active charging/discharging power of the ESS, and voltage magnitude of Bus 6 (before and after mitigation).}
    \vspace*{-2mm}
    \label{combined_profile}
\end{figure}

Assume a combined cyber-physical attack that happens at  12:00, i.e., at time instance 6 that breaks one day into two equal length of periods. 
The cyber resource $x_6$ is attacked and needs to be isolated, and the generator $P_{6,g}^{\mathrm{gen}}$ at bus 6 is assumed to be out of operation caused by a physical disruption. 
Fig. \ref{cyber_network_root6} shows the reconfiguration of a cyber communication network after isolating Node 6. 
% 
% 
% \revfirst{{\bf{REV 1.2}}}{\color{blue} 
A set of neighboring cyber nodes  $\mathcal{M}_{\tilde{6}} = \{11,12,13\}$ are candidates for replacing the cyber resource $x_6$. Following Algorithm \ref{alg_1}, the neighboring cyber Node 11 is selected with minimal costs to reroute the cyber network. Subsequently, Node 11 controls the physical resources of bus 6, such as controlling or starting the ESS at bus 6.
% }
% , and Node 11 is selected with minimal costs to reroute the cyber network. 
% 
% 
Fig. \ref{all_generator_after_attack} shows the 
% \ref{voltage_and_load}
% 
% \begin{figure}[!htb]
% % \vspace*{-2mm}
%     \centering
%     \includegraphics[width=0.45\textwidth, trim = 0mm 0mm 0mm 0mm, clip]{images/voltages_bw.png}
%     % \vspace*{-3mm}
%     \caption{Voltage magnitudes of all buses (blue and red dashed lines denote the lower and upper voltage bounds, respectively).}
%     % \vspace*{-4mm}
%     \label{voltage_and_load}
% \end{figure}
shows the generators' generations after the cyber-physical attack. 
As can be seen, Generator 6 stops functioning after time slot 6 with no active power generation. 
% 
% shows the voltage magnitudes of all buses before and after the cyber-physical attack. All bus voltages are maintained within the voltage limit. 
% 
% 
% 
In Fig. \ref{combined_profile}, 
% 
% \begin{figure}[!htb]
% \vspace*{-2mm}
%     \centering
%     \includegraphics[width=0.45\textwidth, trim = 0mm 0mm 0mm 0mm, clip]{images/combined_profile_2.png}
%     % \vspace*{-3mm}
%     \caption{Active power generation of Generator and ESS charging/discharging profiles (before and after the attack).}
%     \vspace*{-4mm}
%     \label{generators}
% \end{figure}
the upper figure shows that the ESS at Bus 6 starts to power the physical systems after the attack. The lower figure shows that the voltage magnitude of Bus 6 drops quickly after the attack, and the mitigation is effective in regulating the voltage to stay within the voltage limits.

% the upper and lower figures show that the cyber-physical attack led to the breakdown of Generator 6, and the ESS at Bus 6 starts to power the physical systems continuously after the attack. 
% 

\section{Conclusion and Future Work}
\label{Conclusion}

In this paper, we present a synthesized cyber-physical resource optimization framework that cooperatively controls the cyber and physical resources within the power systems.
By formulating the problem as a constrained optimization problem with cyber, physical, and cyber-physical objectives and constraints, we develop bi-level solutions to optimize the operation of cyber communication networks and physical power resources.
The proposed method enhances grid resilience by cooperating the cyber and physical side actions, with adaptive response to cyber-physical contingencies. 
Numerical simulations on the modified IEEE 14-bus system validate the effectiveness of our approach, demonstrating its potential for the integrated control of grid-edge cyber and physical resources.
Future work includes scaling up the solution to the large-scale cyber topology.  

\vspace{-1mm}

% \section*{Acknowledgment}

% The authors would like to acknowledge the
% National Science Foundation under Grant 2220347 and the US Department of Energy under award DE-CR0000018, for supporting this work.

\bibliographystyle{IEEEtran}

\bibliography{bibliography}

\end{document}